\documentstyle[amssymb,preprint,aps]{revtex}

\begin{document}
\title{Glue content and mixing angle of the $\eta -\eta ^{\prime }$ system. The
effect of the isoscalar $0^{-}$ continuum.}
\author{N. F. Nasrallah\thanks{%
e-mail address: nsrallh@ul.edu.lb}}
\address{Faculty of Science, Lebanese University\\
Tripoli, Lebanon}
\maketitle

\begin{abstract}
Masses and topological charges of the $\eta $ and $\eta ^{\prime }$ mesons
are expressed in terms of the singlet-octet mixing angle $\theta $.
Contributions of the pseudoscalar $0^{-}$ continuum are evaluated in a \
model independant way. Applications to the decay $\eta \rightarrow 3\pi $
and to the radiative decay of vector mesons involving $\eta $ and $\eta
^{\prime }$ are considered. Agreement with experiment \ is in general good
and the results quite stable for $-30.5%
{{}^\circ}%
\lesssim \theta \lesssim -18.5%
{{}^\circ}%
$.
\end{abstract}


\section{Introduction}

The subject of $\ \eta -\eta $ mixing has been a topic of discussion from
the time $SU(3)$ flavour symmetry was proposed.The gluon axial anomaly and
the corresponding topoligical charges of the isoscalar mesons imply that the 
$SU(3)$ singlet axial-vector current is not conserved in the chiral limit.
Traditionally,analyses based on the Gell-Mann-Okubo mass formula led to the
adoption of a small value for the octet-singlet \ mixing angle, $\theta
\simeq -10%
{{}^\circ}%
$. Subsequent study of the axial-anomaly generated decays $\eta ,$\ $\eta
^{\prime }\rightarrow 2\gamma $ and measurement of the decay rates led to
adoption of larger values for the mixing angle, $\theta =-25%
{{}^\circ}%
$ - $-20%
{{}^\circ}%
$ \cite{Lin}. Phenomenological applications of the mixing and of the
anomalies were considered \cite{Ball}\cite{feldmann}. In ref. \cite
{nasrallah} the PCAC corrections to the calculation of the decay rates\ of
the $\eta $\ and $\eta ^{\prime }$ to two photons, i.e the corrections to
the soft meson limits arising from the large masses of the isoscalars were
evaluated and found to be quite large.

It is the purpose of the present work is to express the mases and
topological charges of the $\eta $\ and $\eta ^{\prime }$\ mesons in terms
of the mixing angle $\theta .$Contributions of the pseudoscalar $0^{-}$
continuum will be taken care of in a model independant way using only\ known 
${\rm QCD}$ expressions of spectral functions in the deep euclidean region
and the fact that the main contribution to dispersion integrals which enter
in the evaluation of $SU(3)$x$SU(3)$ breaking arise from the energy interval 
$1.5{\rm GeV}^{2}-2.5{\rm GeV}^{2}$ where the isoscalar pseudoscalar
resonances lie.

In section II the quark and gluon content of the $\eta $ and $\eta ^{\prime
} $\ are related to the mixing angle $\theta $. In section III the \ decay
rate $\Gamma $\ $(\eta \rightarrow 3\pi )$ is related to the quark mass
ratio $\frac{m_{u}}{m_{d}}$ and values of the latter infered.In section IV
the analysis of Ball et al. \cite{Ball} of the decays $P\rightarrow V\gamma $
is modified to take chiral symmetry breaking into account.

Agreement with experiment is in general good and the results are quite
insensitive to exact values of the mixing angle $\theta $.

\section{Masses of the $\protect\eta $ and $\protect\eta ^{\prime }$}

The isoscalar components of the octet of axial-vector currents couple to the
physical $\eta $ and $\eta ^{\prime }$ mesons

\begin{eqnarray}
\langle 0\left| A_{\mu }^{(8)}\right| \eta (p)\rangle &=&2if_{8}\cos \theta
p_{\mu }  \nonumber \\
\langle 0\left| A_{\mu }^{(0)}\right| \eta (p)\rangle &=&-2if_{0}\sin \theta
p_{\mu }  \eqnum{2.1}  \label{2.1} \\
\langle 0\left| A_{\mu }^{(8)}\right| \eta ^{\prime }(p)\rangle
&=&2if_{8}\sin \theta p_{\mu }  \nonumber \\
\langle 0\left| A_{\mu }^{(0)}\right| \eta ^{\prime }(p)\rangle
&=&2if_{0}\cos \theta p_{\mu }  \nonumber
\end{eqnarray}
$\theta $ is the singlet-octet mixing angle. In the $SU(3)$ limit $%
f_{8}=f_{\pi }=92.4{\rm MeV}.$ The axial-vector currents are given in terms
of the quark fields 
\begin{eqnarray}
A_{\mu }^{(8)} &=&\frac{1}{\sqrt{3}}\left( \stackrel{-}{u}\gamma _{\mu
}\gamma _{5}u+\stackrel{\_}{d}\gamma _{\mu }\gamma _{5}d-2\stackrel{\_}{s}%
\gamma _{\mu }\gamma _{5}s\right)  \nonumber \\
A_{\mu }^{(0)} &=&\sqrt{\frac{2}{3}}\left( \stackrel{\_}{u}\gamma _{\mu
}\gamma _{5}u+\stackrel{\_}{d}\gamma _{\mu }\gamma _{5}d+\stackrel{\_}{s}%
\gamma _{\mu }\gamma _{5}s\right)  \eqnum{2.2}  \label{2.2}
\end{eqnarray}
Unlike $f_{\pi }$, $f_{0}$ and $f_{8}$ are not related to any physical
process. The currents which project on the physical $\eta \ $and $\eta
^{\prime }$ states are, respectively 
\begin{eqnarray}
A_{\mu }^{(\eta )} &=&\left( \frac{A_{\mu }^{(8)}\cos \theta }{f_{8}}-\frac{%
A_{\mu }^{(0)}\sin \theta }{f_{0}}\right)  \nonumber \\
A_{\mu }^{(\eta ^{\prime })} &=&\left( \frac{A_{\mu }^{(8)}\sin \theta }{%
f_{8}}+\frac{A_{\mu }^{(0)}\cos \theta }{f_{0}}\right)  \eqnum{2.3}
\label{2.3}
\end{eqnarray}
i.e. 
\begin{equation}
\langle 0\left| A_{\mu }^{(\eta )}\right| \eta (p)\rangle =2ip_{\mu }\ \ \
,\ \ \ \ \langle 0\left| A_{\mu }^{(\eta )}\right| \eta ^{\prime }(p)\rangle
=0  \eqnum{2.4}  \label{2.4}
\end{equation}
and 
\begin{equation}
\langle 0\left| A_{\mu }^{(\eta ^{\prime })}\right| \eta (p)\rangle =0\ \ \
\ ,\ \ \ \ \langle 0\left| A_{\mu }^{(\eta ^{\prime })}\right| \eta ^{\prime
}(p)\rangle =2ip_{\mu }  \eqnum{2.5}  \label{2.5}
\end{equation}
When the divergence of the currents is taken, the singlet component picks up
a gluon anomaly term 
\begin{eqnarray}
\partial _{\mu }A_{\mu }^{(8)} &=&\frac{2}{\sqrt{3}}\left( m_{u}\stackrel{\_%
}{u}i\gamma _{5}u+m_{d}\stackrel{\_}{d}i\gamma _{5}d-2m_{s}\stackrel{\_}{s}%
i\gamma _{5}s\right)  \nonumber \\
\partial _{\mu }A_{\mu }^{(0)} &=&2\sqrt{\frac{2}{3}}\left( m_{u}\stackrel{\_%
}{u}i\gamma _{5}u+m_{d}\stackrel{\_}{d}i\gamma _{5}d+m_{s}\stackrel{\_}{s}%
i\gamma _{5}s\right) +\sqrt{\frac{2}{3}}\frac{3\alpha _{s}}{4\pi }G%
\widetilde{G}  \eqnum{2.6}  \label{2.6}
\end{eqnarray}
With $G\widetilde{G}=G_{\mu \nu }\widetilde{G}^{\mu \nu },G_{\mu \nu }$
being the gluonic field strength tensor and $\widetilde{G}_{\mu \nu }=\frac{1%
}{2}\epsilon _{\mu \nu \rho \sigma }G^{\rho \sigma }$ it's dual. As $%
m_{u,d}\ll m_{s}$, 
\begin{equation}
\partial _{\mu }A_{\mu }^{(8)}\backsimeq \frac{-4}{\sqrt{3}}m_{s}\stackrel{\_%
}{s}i\gamma _{5}s  \eqnum{2.7}  \label{2.7}
\end{equation}
\begin{equation}
\partial _{\mu }A_{\mu }^{(0)}\backsimeq 2\sqrt{\frac{2}{3}}m_{s}\stackrel{\_%
}{s}i\gamma _{5}s+\sqrt{\frac{2}{3}}\frac{3\alpha _{s}}{4\pi }G\widetilde{G}
\eqnum{2.8}  \label{2.8}
\end{equation}
constitute good approximations. There results for the masses of the $\eta $
and $\eta ^{\prime }$%
\begin{equation}
m_{\eta }^{2}=\frac{1}{\sqrt{3}f_{\pi }}\left( \frac{\cos \theta }{F_{8}}+%
\frac{\sin \theta }{\sqrt{2}f_{0}}\right) \langle 0\left| -2im_{s}\stackrel{%
\_}{s}\gamma _{5}s\right| \eta \rangle -\frac{\sin \theta }{\sqrt{6}f_{\pi
}F_{0}}\langle 0\left| \frac{3\alpha _{s}}{4\pi }G\widetilde{G}\right| \eta
\rangle  \eqnum{2.9}  \label{2.9}
\end{equation}
\begin{equation}
m_{\eta ^{\prime }}^{2}=\frac{1}{\sqrt{3}f_{\pi }}\left( \frac{\sin \theta }{%
F_{8}}-\frac{\cos \theta }{\sqrt{2}f_{0}}\right) \langle 0\left| -2im_{s}%
\stackrel{\_}{s}\gamma _{5}s\right| \ \eta ^{\prime }\rangle +\frac{\cos
\theta }{\sqrt{6}f_{\pi }F_{0}}\langle 0\left| \frac{3\alpha _{s}}{4\pi }G%
\widetilde{G}\right| \eta ^{\prime }\rangle  \eqnum{2.10}  \label{2.10}
\end{equation}
with the notation \ $f_{0,8}=F_{0,8}f_{\pi }$. Using eq.(\ref{2.5}) one has
for the topological charges

\begin{eqnarray}
\langle 0\left| \frac{3\alpha _{s}}{4\pi }G\widetilde{G}\right| \eta \rangle
&=&\sqrt{3}f_{\pi }m_{\eta }^{2}\left( F_{8}\cos \theta -\sqrt{2}F_{0}\sin
\theta \right)  \nonumber \\
\langle 0\left| \frac{3\alpha _{s}}{4\pi }G\widetilde{G}\right| \eta
^{\prime }\rangle &=&\sqrt{3}f_{\pi }m_{\eta ^{\prime }}^{2}\left( F_{8}\sin
\theta +\sqrt{2}F_{0}\cos \theta \right)  \eqnum{2.11}  \label{2.11}
\end{eqnarray}
The next step consists in the evaluation of the matrix elements $\langle
0\left| -2im_{s}\stackrel{\_}{s}\gamma _{5}s\right| \eta ,\eta ^{\prime
}\rangle .$ Consider 
\begin{equation}
\Pi _{\mu \nu }(q^{2})=i\int dxe^{iqx}\langle 0\left| TA_{\mu }^{(\eta
)}(x)A_{\nu }^{(\eta )^{\dagger }}(0)\right| 0\rangle =(q_{\mu }q_{\nu
}-q^{2}g_{\mu \nu })\Pi _{1}(q^{2})+q_{\mu }q_{\nu }\Pi _{2}(q^{2}) 
\eqnum{2.12}  \label{2.12}
\end{equation}
The Ward-Takahashi identity gives 
\begin{equation}
-iq_{\mu }\Pi _{\mu \nu }=i\int dxe^{iqx}\langle 0\left| T\partial _{\mu
}A_{\mu }^{(\eta )}(x)A_{\nu }^{(\eta )^{\dagger }}(0)\right| 0\rangle
=-iq^{2}q_{\nu }\Pi _{0}(q^{2})  \eqnum{2.13}  \label{2.13}
\end{equation}
Defining\ $T(q^{2})=-iq^{2}\Pi _{0}(q^{2})$ and isolating the $\eta $ pole 
\begin{equation}
T(q^{2})=-2i\frac{\langle 0\left| \partial _{_{\mu }}A_{\mu }^{(\mu
)}\right| \eta \rangle }{(q^{2}-m_{\eta }^{2})}+...  \eqnum{2.14}
\label{2.14}
\end{equation}
In the deep euclidean region the behaviour of $T(q^{2)}$ is known \cite{wu} 
\begin{equation}
T^{{\rm QCD}}(q^{2})=\frac{8}{3}\left( \frac{\sqrt{2}\cos \theta }{f_{8}}+%
\frac{\sin \theta }{f_{0}}\right) ^{2}\frac{\langle 0\left| -m_{s}\stackrel{%
\_}{s}s\right| 0\rangle }{q^{2}}+...  \eqnum{2.15}  \label{2.15}
\end{equation}
Define $F(t)=tT(t)$ with \ $t=(q^{2}-m_{\eta }^{2}).$ $F(t)$ is an analytic
function in the complex $t$ plane with a cut starting at $t=(m_{\eta
}+2m_{\pi })^{2}-m_{\eta }^{2}$ and running along the positive real axis. As
a result of Cauchy's theorem then

\begin{equation}
F(0)=-2i\langle 0\left| \partial _{\eta }A_{\mu }^{(\eta )}\right| \eta
\rangle =\frac{1}{2\pi i}\int_{c}\frac{dt}{t}F(t)  \eqnum{2.16}  \label{2.16}
\end{equation}
where $c$ is a closed contour consisting of a circle of large radius and two
straight lines parallel to the x-axis just above and just below the cut. The
main contribution to the integral above is provided by the interval $1.5{\rm %
GeV}^{2}\lesssim q^{2}\lesssim 2.5{\rm GeV}^{2}$which includes the two
pseudoscalar excitations $\eta (1295)$ and $\eta (1440)$. These resonances
couple to the $\eta $ and $\eta ^{\prime }$with unknown strengths. On the
circle $F(t)\simeq $ $F^{{\rm QCD}}(t)$ except for a small region near the
real axis. Instead of expression (\ref{2.16}) consider the following
modified integral

\begin{equation}
F(0)=\frac{1}{2\pi i}\int_{c}dt\left( \frac{1}{t}-a_{0}-a_{1}t\right) F(t) 
\eqnum{2.17}  \label{2.17}
\end{equation}
The coefficients $a_{0}$ and $a_{1}$ are chosen so as to annihilate the
integrand at $q^{2}=m_{1}^{2}=1.66$ ${\rm GeV}^{2}$ and at $%
q^{2}=m_{2}^{2}=2.04{\rm GeV}^{2}$%
\begin{equation}
a_{0}=\frac{1}{\mu _{1}^{2}}+\frac{1}{\mu _{2}^{2}}\ \ \ ,\ \ \ a_{1}=-\frac{%
1}{\mu _{1}^{2}\mu _{2}^{2}}\ \ \ ,\mu _{1,2}^{2}=m_{1,2}^{2}-m_{\eta }^{2} 
\eqnum{2.18}  \label{2.18}
\end{equation}
The contribution of the isoscalar pseudoscalar resonances is thus practicaly
eliminated and the integrand is \ reduced to only a few percent of it's
initial value over the interval $1.5{\rm GeV}^{2}\lesssim q^{2}\lesssim 2.5%
{\rm GeV}^{2}.$

The main contribution to the integral (\ref{2.17}) comes now from the circle 
\begin{equation}
F(0)\backsimeq \frac{1}{2\pi i}\oint dt\left( \frac{1}{t}-a_{0}-a_{1}t%
\right) F^{{\rm QCD}}(t)  \eqnum{2.19}  \label{2.19}
\end{equation}
So that using eq.(\ref{2.15}) 
\begin{equation}
\langle 0\left| -2im_{s}\stackrel{\_}{s}\gamma _{5}s\right| \eta \rangle =2i%
\sqrt{\frac{2}{3}}\left( \frac{\sqrt{2}\cos \theta }{f_{8}}+\frac{\sin
\theta }{f_{0}}\right) \left( 1+\frac{m_{\eta }^{2}}{\mu _{1}^{2}}+\frac{%
m_{\eta }^{2}}{\mu _{2}^{2}}+\frac{m_{\eta }^{4}}{\mu _{1}^{2}\mu _{2}^{2}}%
\right) \langle 0\left| -m_{s}\stackrel{\_}{s}s\right| 0\rangle  \eqnum{2.20}
\label{2.20}
\end{equation}
In the equation above the first bracket shows the effect of mixing the
second one represents the correction factor due to the isoscalar $0^{-}$
continuum,numerically it amounts to $1.431$ .In a similar fashion one obtains

\begin{equation}
\langle 0\left| 2im_{s}\stackrel{\_}{s}\gamma _{5}s\right| \eta ^{\prime
}\rangle =2i\sqrt{\frac{2}{3}}\left( \frac{\cos \theta }{f_{0}}-\frac{\sqrt{2%
}\sin \theta }{f_{8}}\right) \left( 1+\frac{m_{\eta ^{\prime }}^{2}}{\mu
_{1}^{^{\prime }2}}+\frac{m_{\eta ^{\prime }}^{2}}{\mu _{2}^{\prime 2}}+%
\frac{m_{\eta ^{\prime }}^{4}}{\mu _{1^{\prime }}^{2}\mu _{2^{\prime }}^{2}}%
\right) \langle 0\left| -m_{s}\stackrel{\_}{s}s\right| 0\rangle 
\eqnum{2.21}  \label{2.21}
\end{equation}
The correction factor due to the continuum now amounts to $4.090$ . If the
Gell-Mann-Oakes-Renner expression $\langle 0\left| -m_{s}\stackrel{\_}{s}%
\gamma _{5}s\right| 0\rangle =f_{K}^{2}m_{K}^{2}$ is used ,the masses of the 
$\eta $ and $\eta ^{\prime }$ mesons come out

\begin{equation}
m_{\eta }^{2}=\left[ .352\left( \frac{\sqrt{2}\cos \theta }{F_{8}}+\frac{%
\sin \theta }{F_{0}}\right) -.212\sin \theta F_{8}\left( \frac{\cos \theta }{%
F_{0}}-\frac{\sqrt{2}\sin \theta }{F_{8}}\right) \right] {\rm GeV}^{2} 
\eqnum{2.22}  \label{2.22}
\end{equation}
\begin{equation}
m_{\eta ^{\prime }}^{2}=\left[ 1.006\left( \frac{\cos \theta }{F_{0}}-\frac{%
\sqrt{2}\sin \theta }{F_{8}}\right) +.649\cos \theta F_{8}\left( \frac{\sqrt{%
2}\cos \theta }{F_{8}}+\frac{\sin \theta }{F_{0}}\right) \right] {\rm GeV}%
^{2}  \eqnum{2.23}  \label{2.23}
\end{equation}
The unknowns $\theta $ ,$F_{0}$ and $F_{8}$ also appear in the expressions
of the widths of the anomaly generated two photon decays of the $\eta $ and $%
\eta ^{\prime }$ \cite{Adler}.

\begin{equation}
\Gamma (\eta \rightarrow 2\gamma )=\frac{\alpha ^{2}m_{\eta }^{3}}{192\pi
^{3}f_{\pi }^{2}}\left( \frac{\cos \theta }{F_{8}}-\frac{2\sqrt{2}\sin
\theta }{F_{0}}\right) ^{2}(1+\Delta _{\eta })^{2}=(.465\pm .043){\rm keV} 
\eqnum{2.24}  \label{2.24}
\end{equation}
\begin{equation}
\Gamma (\eta ^{\prime }\rightarrow 2\gamma )=\frac{^{\alpha ^{2}m_{\eta
^{\prime }}^{3}}}{192\pi ^{3}f_{\pi }^{2}}\left( \frac{\sin \theta }{F_{8}}+%
\frac{2\sqrt{2}\cos \theta }{F_{0}}\right) (1+\Delta _{\eta ^{\prime
}})^{2}=(4.28\pm .64){\rm keV}  \eqnum{2.25}  \label{2.25}
\end{equation}
$\Delta _{\eta ,\eta ^{\prime }}$ result from the chiral symmetry breaking
corrections, i.e they represent the deviations from the soft meson limits
arising from the large masses of the $\eta $ and $\eta ^{\prime }$. They
were evaluated in \cite{nasrallah}and found to be quite large, $\Delta
_{\eta }=.77$ and $\Delta _{\eta ^{\prime }}=6.0.$ Numerically, then

\begin{equation}
\frac{\cos \theta }{F_{8}}-\frac{2\sqrt{2}\sin \theta }{F_{0}}=.929\pm .094 
\eqnum{2.26}  \label{2.26}
\end{equation}
\begin{equation}
\frac{\sin \theta }{F_{8}}+\frac{2\sqrt{2}\cos \theta }{F_{0}}=.308\pm .050 
\eqnum{2.27}  \label{2.27}
\end{equation}
The errors in the equations above represent only the experimental
uncertainties in the decay rates. Four equations (\ref{2.22}), (\ref{2.23}),
(\ref{2.24}) and (\ref{2.25})\ are now available. The straightforward
procedure is to vary $\theta $, use eqs. (\ref{2.26}) and (\ref{2.27}) and
see what eqs. (\ref{2.22}) and (\ref{2.23})\ give for the masses. Two
limiting cases are considered

\begin{equation}
\theta =-18.5%
{{}^\circ}%
:\ \ \ \ \ \ \ \ m_{\eta }=.547{\rm GeV}\ \ \ \ \ \ ,\ \ m_{\eta ^{\prime
}}=1.011{\rm GeV}  \eqnum{2.28}  \label{2.28}
\end{equation}
$m_{\eta }$ is adjusted to it's experimental value, $m_{\eta ^{\prime }}$%
deviates by $5.5\%.$

\begin{equation}
\theta =-30.5%
{{}^\circ}%
:\ \ \ \ \ \ \ m_{\eta }=.468{\rm GeV}\ \ \ \ \ \ \ ,\ m_{\eta ^{\prime
}}=.958{\rm GeV}  \eqnum{2.29}  \label{2.29}
\end{equation}
$m_{\eta ^{\prime }}$ is adjusted to it's experimental value, $m_{\eta }$
deviates by $14\%.$

Because of the approximations made (vector meson dominance in obtaining $%
\Delta _{\eta }$ and $\Delta _{\eta ^{\prime }}$ in \cite{nasrallah}\ etc.)
it is seen that both set of values (\ref{2.28}) and (\ref{2.29})\ fall well
within the expected range. This applies as well to all values of the masses
resulting from values of $\theta $ between the limits given above. This
stability of course forbids any prediction for the exact value of the mixing
angle. We shall see that the same applies to all predictions of the present
work, i.e. they show a remarkable stability against variations of the mixing
angle. The next topic of investigation is the decay $\eta \rightarrow 3\pi .$

\section{$\protect\eta \rightarrow 3\protect\pi $}

Consider the neutral mode $\eta \rightarrow 3\pi 
{{}^\circ}%
.$ As is well known this decay proceeds through the isospin breaking part of
the {\rm QCD} Hamiltonian 
\begin{equation}
H=\frac{1}{2}(m_{u}-m_{d})(\stackrel{\_}{u}u-\stackrel{\_}{d}d)  \eqnum{3.1}
\label{3.1}
\end{equation}
Standard Current -Algebra soft pion techniques yield for the decay rate\cite
{Ball} 
\begin{equation}
\Gamma (\eta \rightarrow 3\pi 
{{}^\circ}%
)=\frac{\sqrt{3}}{6912\pi ^{2}}\delta _{\eta }\frac{(m_{\eta }-3m_{\pi })^{2}%
}{m_{\eta }}\frac{(m_{d}-m_{u)^{2}}}{f_{\pi }^{6}}\left| \langle 0\left| 
\stackrel{\_}{u}\gamma _{5}u\right| \eta \rangle \right| ^{2}  \eqnum{3.2}
\label{3.2}
\end{equation}
$\delta _{\eta }=.86$ is a kinematical factor.

The matrix element $\langle 0\left| \stackrel{\_}{u}\gamma _{5}u\right| \eta
\rangle $ can be evaluated in exactly the same way $\langle 0\left| 
\stackrel{\_}{s}\gamma _{5}s\right| \eta \rangle $ was with the result 
\begin{equation}
\langle 0\left| \stackrel{\_}{u}\gamma _{5}u\right| \eta \rangle =\frac{-1}{%
\sqrt{3}f_{\pi }}\left( \frac{\cos \theta }{F_{8}}-\frac{\sqrt{2}\sin \theta 
}{F_{0}}\right) \left( 1+\frac{m_{\eta }^{2}}{\mu _{1}^{2}}+\frac{m_{\eta
}^{2}}{\mu _{2}^{2}}+\frac{m_{\eta }^{4}}{\mu _{1}^{2}\mu _{2}^{2}}\right)
\langle 0\left| \stackrel{\_}{u}u\right| 0\rangle  \eqnum{3.3}  \label{3.3}
\end{equation}
Note that the continuum enhancement factor appearing in the second bracket
is the same as in eq.(\ref{2.20} ) as it arises from the same isoscalar $0^{-}$
continuum\ . The Gell-Mann-Oakes -Renner expression $\langle 0\left|
-(m_{u}+m_{d})\stackrel{\_}{u}u\right| 0\rangle =f_{\pi }^{2}m_{\pi }^{2}$
gives then for the decay rate

\begin{equation}
\Gamma (\eta \rightarrow 3\pi 
{{}^\circ}%
)=\frac{1.431^{2}}{\sqrt{3}6912\pi ^{2}}\left( \frac{\cos \theta }{F_{8}}-%
\frac{\sqrt{2}\sin \theta }{F_{0}}\right) ^{2}\delta _{\eta }\frac{(m_{\eta
}-3m_{\pi })^{2}}{m_{\eta }}\left( \frac{m_{d}-m_{u}}{m_{d}+m_{u}}\right)
^{2}\frac{m_{\pi }^{4}}{f_{\pi }^{4}}(1+C)  \eqnum{3.4}  \label{3.4}
\end{equation}
$C$ is a rescattering enhancement factor which was estimated to be $.83$ by
Roiesnel and Truong \cite{truong}. It was argued however by Gasser and
Leutwyler \cite{Ball} that one should take instead $C\simeq .50$ a value we
shall here use. Equation (\ref{3.4}) can now be used to evaluate the ratio $%
\frac{m_{u}}{m_{d}}$ of the quark masses. For $\theta =-18.5%
{{}^\circ}%
$ one gets $\frac{m_{u}}{m_{d}}=.47$. This ratio decreases to$\frac{m_{u}}{%
m_{d}}=.42$ for $\theta =-30.5%
{{}^\circ}%
.$

\section{Radiative vector-meson decays}

The radiative decays of vector-mesons involving $\eta $ or $\eta ^{\prime }$
were studied in \cite{Ball} using the hypothesis of vector-meson dominance
to relate them to the two photon decays of $\eta $ and $\eta ^{\prime }.$
The only improvement we can provide is to multiply the expressions of Ball
et al$.$ \cite{Ball} for the coupling constants by the chiral symmetry
breaking enhancement factors $(1+\Delta _{\eta ,\eta ^{\prime }})$, thus 
\begin{equation}
g_{\rho \eta \gamma }=\frac{1}{4\pi }\sqrt{\frac{3}{2}}\frac{m_{\rho }}{%
f_{\rho }\pi ^{2}}\left( \frac{\cos \theta }{F_{8}}-\frac{\sqrt{2}\sin
\theta }{F_{0}}\right) (1+\Delta _{\eta })  \eqnum{4.1}  \label{4.1}
\end{equation}
\begin{equation}
g_{\rho \eta ^{\prime }\gamma }=\frac{1}{4\pi }\sqrt{\frac{3}{2}}\frac{%
m_{\rho }}{f_{\rho }\pi ^{2}}\left( \frac{\sin \theta }{F_{8}}+\frac{\sqrt{2}%
\cos \theta }{F_{0}}\right) (1+\Delta _{\eta ^{\prime }})  \eqnum{4.2}
\label{4.2}
\end{equation}
etc.

Decays involving $\omega $ and $\phi $ vector-mesons suffer from an
additional uncertainty in the $\omega -\phi $ mixing angle to which the
decay rates might be very sensitive. The results for $g_{VP\gamma }$ in {\rm %
GeV}$^{-1}$ are listed in table 1

\bigskip

$%
\mathrel{\mathop{%
\begin{tabular}{|l|l|l|l|l|l|l|}
\hline
 & $g_{\rho \eta \gamma }$ & $g_{\rho \eta ^{\prime }\gamma }$ & $g_{\omega \eta \gamma }$ & $g_{\omega \eta ^{\prime }\gamma }$ & $g_{\phi \eta \gamma }$ & $g_{\phi \eta ^{\prime }\gamma }$ \\ \hline
Experiment & $1.85\pm .35$ & $\lesssim 1.31$ & $.60\pm .15$ & $.45\pm .06$ & $.70\pm .03$ & $.73\pm .16$ \\ \hline
$\theta =-18.5%
{{}^\circ}$ & $1.57$ & $.59$ & $.52$ & $.56$ & $1.06$ & $2.43$ \\ \hline
$\theta =-30.5%
{{}^\circ}$ & $1.38$ & $.35$ & $.50$ & $.52$ & $.68$ & $2.93$ \\ \hline
\end{tabular}
}\limits_{Table\;1}}%
$

\bigskip

Once again the results are quite stable against variations of the mixing
angle and agreement with experiment is reasonnably good except for $g_{\phi
\eta ^{\prime }\gamma }.$

An additional item of investigation is the ratio $r=\frac{\langle 0\left| G%
\widetilde{G}\right| \eta ^{\prime }\rangle }{\langle 0\left| G\widetilde{G}%
\right| \eta \rangle }$\ for which we get

\begin{equation}
r=5.1\ \ \ \ \ \ \ \ for\ \ \ \ \ \theta =-18.5%
{{}^\circ}%
\eqnum{4.3}  \label{4.3}
\end{equation}
and 
\begin{equation}
r=2.64\ \ \ \ \ \ for\ \ \ \ \ \theta =-30.5%
{{}^\circ}%
\eqnum{4.4}  \label{4.4}
\end{equation}

Novikov et al. \cite{Novikov}, assuming that the decay mechanism for $J/\Psi
\rightarrow \eta (\eta ^{\prime })\gamma $ via the strong anomaly dominates
in addition to vector meson dominance in the $J/\Psi $ channel, relate $r$
to the ratio of the corresponding decay rates, they give

\begin{equation}
r=\sqrt{\frac{\Gamma (J/\Psi \rightarrow \eta ^{\prime }\gamma )}{\Gamma
(J/\Psi \rightarrow \eta \gamma )}}\left( \frac{1-\frac{m_{\eta }^{2}}{%
m_{J/\Psi }^{2}}}{1--\frac{m_{\eta ^{\prime }}^{2}}{m_{J/\Psi }^{2}}}\right)
^{\frac{3}{2}}=2.48\pm .15  \eqnum{4.5}  \label{4.5}
\end{equation}
which is reasonnably consistent with our values.

\section{Conclusion}

The effect of the contribution of the isoscalar pseudoscalar continuum on
the glue content and mixing angle of the $\eta -\eta ^{\prime }$ system. The
masses of the isoscalar mesons and their decay rates into two photons are
consistently described in terms of a single octet-singlet mixing angle $%
\theta $ and their values are quite stable against variations of $\theta .$
Applications to the decays $\eta \rightarrow 3\pi $ and $V\rightarrow
P\gamma $ were also considered. Agreement with experiment is in general
quite good.

\end{document}